\documentclass[a4paper,12pt,notitlepage]{article}

\usepackage[latin1]{inputenc} 
\usepackage[T1]{fontenc}
\usepackage{graphicx}
\usepackage{natbib}
\usepackage{amsmath,amssymb}

\begin{document}

\title{A Dynamical Approach to Operational Risk Measurement}
\author{Marco Bardoscia\footnote{corresponding author: \texttt{marco.bardoscia@ba.infn.it}}, Roberto Bellotti \\
		\\
		\small{\emph{Universit\`{a} degli Studi di Bari - Dipartimento di Fisica ``M.Merlin''}} \\
		\small{\emph{Istituto Nazionale di Fisica Nucelare, Sezione di Bari}}}

\date{ }

\maketitle

\begin{abstract}
We propose a dynamical model for the estimation of Operational Risk in banking institutions. Operational Risk is the risk that a financial loss occurs as the result of failed processes. Examples of operational losses are the ones generated by internal frauds, human errors or failed transactions. In order to encompass the most heterogeneous set of processes, in our approach the losses of each process are generated by the interplay among random noise, interactions with other processes and the efforts the bank makes to avoid losses. We show how some relevant parameters of the model can be estimated from a database of historical operational losses, validate the estimation procedure and test the forecasting power of the model. Some advantages of our approach over the traditional statistical techniques are that it allows to follow the whole time evolution of the losses and to take into account different-time correlations among the processes.
\end{abstract}

\section{Introduction} \label{sec:intro}
Operational Risk (OR) is defined as ``the risk of [financial] loss resulting from inadequate or failed internal processes, people and systems or from external events'', including legal risk, but excluding strategic and reputation linked risks. The previous definition is taken from The New Basel Capital Accord or Basel II \citep{basel}, which can be viewed as the turning point for the OR management within banking institutions. In fact Basel II considers OR to be a critical risk factor. The accord requires banks to cope with OR by setting a side a certain capital charge, and it proposes three methods for calculating it.

The first method is the Basic Indicator Approach (BIA) that sets that capital to the $15\%$ of the bank's gross income. The second is the STandardized Approach (STA), which is a simple generalization of the BIA: the percentage of the gross income is different for each business line and goes from $12\%$ to $18\%$. The BIA and the STA share the fundamental drawback that, because of their inherent assumption that the capital charge is only a linear function of the bank's gross income (per business line), they seem not to be solidly founded. Moreover, they give no additional insight into the dynamics governing the production of operational losses and thus they do not indicate any strategies for lowering them: in other words they do not offer a path toward OR management \citep{mcneil-frey-embrechts, cruz}.

The third method is the Advanced Measurement Approach (AMA) which allows each bank to develop a model of its own that has to satisfy certain requirements and has to be approved by the national regulatory institutions. In particular, all the AMAs are required to use a historical database of internal and external losses. However, because the interest in OR is recent, and because only small, and sometimes unreliable databases of operational losses exist, the assessments of domain experts must also be considered. In addition, every AMA has to respect the classification of losses in eight business lines (like the STAs) and by seven different event types. Usually all the AMAs identify the capital charge with the Value-at-Risk (VaR) over the time horizon of one year and with a confidence level of $99.9\%$, defined as the maximum potential loss not to be exceeded in one year with a confidence level of $99.9\%$, i.e. the 99.9 percentile of the loss distribution. The VaR has a straightforward interpretation: the probability of registering a loss that is higher than the VaR in one year is equal to 0.001 and thus such a loss is registered on average every 1000 years.

Probably the most widely used AMA is the Loss Distribution Approach (LDA), which calculates the loss distribution separately, modeling the distribution of the number of losses over a certain time horizon (frequency) and the distribution of the impact of a single loss (severity) for each couple (business line, event type). The LDA relies on two assumptions:  that frequency and severity for each couple are independent and that the frequency and severity of a couple are independent of the frequency and severity of all the other couples. From this point of view, the main defect of the LDA is that it completely neglects the correlations that may exist among the different couples. Advanced statistical tools aiming at considering the correlations have been proposed \citep{gourier-farkas-abbate, neil-fenton-tailor, cowell-verral-yoon, cornalba-giudici}, but there is still no general consensus regarding their effectiveness.

We believe that there are several advantages in the use of a dynamical model rather than a purely statistical approach such as the LDA. The first one is methodological: in fact, in the context of a purely statistical approach, one has to make assumptions about the shape of a distribution. For the LDA, in particular, one has to specify the frequency and severity distributions before their parameters can be fitted. Also, in the case in which the correlations are modeled through copulae, the functional form of the copula is usually specified a priori. On the other hand, a dynamical model only makes assumptions about the mechanisms underlying the generation of losses and from those assumptions is able to derive the loss distributions. This also means that the basic features of the loss distributions cannot be inserted ``by han'' as in the LDA, but they must emerge from the mechanisms that generate the losses instead. The second advantage is that a dynamical model may account for the different-time correlations. As mentioned previously, much effort has been devoted to including the correlations among different couples in the framework of the LDA. However, since both frequency and severity distributions do not depend on time, it is not possible to deal with the different-time correlations. Let us provide an example to show why, in the context of OR, different-time correlations cannot be neglected. Let us suppose, for example, that a failure occurs in the process of machinery servicing at time $t_1$ because of a damage in the transaction control system that is repaired only later at the time $t_2$. As a consequence, some transactions may fail or be wrongly authorized and generate losses in other processes in the whole time interval $[t_1, t_2]$. As it will be shown throughout the paper, the third advantage is that, in contrast with static approaches, a dynamical model offers a natural and solid framework to forecast future operational losses.

The model proposed by \citet{kuhn-neu, anand-kuhn} is the first attempt to introduce a dynamical model in the context of OR. It can be considered a mixed model in the sense that, while using a dynamical model to derive the frequency distribution, it still uses a static severity distribution. As a consequence, all the aforementioned benefits of adopting a dynamical model are limited to the frequency distribution. The model that we are proposing is completely dynamical in the sense that the relevant variable is the amount of loss registered in each couple at a certain time, avoiding the description in terms of frequency and severity. Indeed, frequency and severity distributions are simply a statistical tool for calculating the loss distributions and, while effective in a static approach, it is difficult to claim that the mechanisms generating losses directly involve them. The equation of motion includes two different mechanisms accounting for the generation of operational losses: the spontaneous generation via a random noise and the generation resulting from the interaction between different couples. The possibility that the bank will invest some money to avoid the occurrence of losses is also taken into account.

The aim of this paper is to set a new viable framework for a dynamical approach to OR. This means that we will try to build a basic dynamical model and find out how its crucial advantages over the purely statistical approaches can be exploited. In particular, it will be shown how the parameters of the model can be estimated from a database of operational losses or assessed by domain experts so that the model can be conveniently tailored to a specific bank. Since the main reason to study a dynamical approach is to try to understand whether, at least in principle, it is able to forecast future operational losses, we test the forecasting power of the model by choosing a realistic set of parameters. As we have already noted, using a dynamical approach implies that one cannot easily impose specific features on the loss distributions a priori: in particular, one cannot rely on the severity distribution to induce heavy tails in the loss distribution, as is the case in \citet{kuhn-neu}. After the preliminary exploration of the potential of the model that we perform in this paper, the next step should be to establish some solid connections between the regions of the parameter space and the desirable features of the loss distributions. We will merely hint at the important role played by the random noise, like some recent papers \citep{bardoscia-bellotti, bardoscia} suggest.

We conclude with a final remark regarding the degrees of freedom of the model. It is very unlikely that the actual dynamics that causes the occurrence of operational losses in a bank directly involves the 56 couples indicated by Basel II: this would imply that the relevant variables for the dynamics of production of operational losses were the same for all the banks, regardless of their organizational structure. For this reason we prefer to introduce a further level of abstraction and just call processes the degrees of freedom of our model, implicitly assuming that they strongly vary from bank to bank and, therefore, that they should be carefully identified by a pool of experts in the internal structure of the bank.

\section{Explaining the model} \label{sec:model}
In the proposed model we associate with each process a positive real valued variable $l_i(t)$ representing the amount of monetary loss incurred at the time $t$ in the $i$-th process. There is one strong motivation to force $l_i(t)$ to assume only positive values: our aim is to learn (at least) some of the parameters of the model from a database of historical operational losses collected by a bank whose entries are all positive, i.e. the quantities that are meant to be observed are intrinsically positive. We point out that negative values of $l_i(t)$ could be interpreted as temporary reserves of money put aside to automatically lower the future losses; forcing $l_i(t)$ to assume only positive values we are excluding this possibility. The evolution of the variables is governed by the following discrete time equation of motion:
\begin{equation} \label{eq:motion}
	l_i(t) = \text{Ramp} \left( \sum_{j=1}^N J_{ij} C_{ij}(t) + \theta_i + \xi_i(t) \right) \, ,
\end{equation}
where $N$ is the number of processes and the ramp function:
\begin{equation*}
	\text{Ramp}(x) = 
	\begin{cases}
		x & \text{for} \;\; x > 0 \\
		0 & \text{for} \;\; x \leq 0
	\end{cases}
\end{equation*}
ensures that $l_i(t)$ remains positive for all the time steps. From eq.\ \ref{eq:motion} we see that the positive terms in the argument of the ramp function aim to generate a loss, while the negative terms aim to avoid the occurrence of a loss. $C_{ij}(t)$ is the number of nonzero losses occurred in the $i$-th process in the
time interval $[t - t_{ij}^*, t - 1]$:
\begin{equation} \label{eq:trigger}
	C_{ij}(t) = \sum_{1 \leq s \leq t_{ij}^*} \Theta \left[ l_j(t-s) \right],
\end{equation}
where $\Theta$ is the Heaviside function and $t_{ij}^*$ is an integer, so that $C_{ij}(t)$ ranges from 0 to $t_{ij}^*$. Eq.\ \ref{eq:trigger} also contains a clear definition of the parameter $t_{ij}^*$. In fact $l_i(t)$ depends on $l_j(t-1), \ldots, l_j(t-t_{ij}^*)$, meaning that $t_{ij}^*$ is the maximum delay up to which the losses incurred in the $j$-th process may influence the losses in the $i$-th process. In this sense the parameter $t_{ij}^*$ is interpreted as the maximum correlation time between the losses incurred in the $i$-th and the $j$-th process; however it should be kept in mind that those correlation times (and thus $t_{ij}^*$) are not symmetrical in general.

Let us now comment the role of each term in the argument of the ramp function in eq.\ \ref{eq:motion}. The sum term in eq.\ \ref{eq:motion} accounts for the potential generation of losses due to the interactions with other processes: if $J_{ij} \neq 0$, then the $i$-th process is influenced by the $j$-th process and, in particular, if $J_{ij} > 0$, each loss occurred in the time interval $[t - t_{ij}^*, t - 1]$ (counted by $C_{ij}(t)$) generates a loss of amount $J_{ij} > 0$ in the $i$-th process at the time $t$. As an example, let us suppose that $J_{12} > 0$ and that $J_{1j} = 0$, $\forall j \neq 2$; from Eqs.\ \ref{eq:motion} and \ref{eq:trigger} we see that the sum term for the first process may assume the values $0, J_{12}, 2 J_{12}, \ldots, t_{ij}^* J_{12}$ depending on the number of nonzero losses that occur in the second process in the time interval $[t - t_{12}^*, t-1]$. This means that each loss that occurs in the second process in this time interval generates a potential loss of amount $J_{12}$ in the first process. As previously explained, $t_{12}^*$	is the maximum correlation time between the losses incurred in the second and in the first process. The losses incurred in the second process before the time $t - t_{12}^*$ have no influence on what may happen to the first process at the time $t$. Obviously this is also true for the losses occurred after the time $t - 1$, i.e. in the future with respect to the time $t$. If the condition $J_{1j} = 0$, $\forall j \neq 2$ does not hold, the effect of the other processes must be summed to the effect of the second process. Also the $J_{ij}$ are not symmetrical in general: in fact the effect on the $i$-th process of a loss incurred in the $j$-th process is clearly different from the effect on the $j$-th process of a loss incurred in the $i$-th process.

In this context, the parameters $t_{ij}^*$ play a crucial role in accounting for different-time correlations and extending the model proposed in \citet{anand-kuhn}, where the status of the $i$-th process depends only on the value of the variables at the time $t - 1$. If the length of the time step of the model is larger than the maximum correlation time among the losses incurred in two different processes, eq.\ \ref{eq:motion} provides a much more realistic description than the equation of motion proposed in \citet{anand-kuhn}. In such a context the length of a time step is the typical time scale of the mechanisms responsible for generating the losses and, as it will be shown in section \ref{sec:parameters_est}, it is linked to the temporal resolution of the database of operational losses that is used to estimate the parameters of the model.

$\theta_i$ can have two very different interpretations depending on its sign. If $\theta_i < 0$, it can be interpreted as the investment the bank makes to avoid the occurrence of losses in the $i$-th process: the greater the value of $|\theta_i|$, the less likely it is that a potential loss due to the other terms becomes an actual loss. Since $\theta_i$ does not depend on time, the amount of money (per unit of time) the bank chooses to invest on each process is fixed a priori rather than dynamically tuned. If $\theta_i > 0$, it can interpreted as pathological tendency of the $i$-th process to produce a loss of amount $\theta_i$ at each time step.

$\xi_i(t)$ is a $\delta$-correlated random noise that accounts for spontaneously generated losses, i.e. losses not caused by the interactions with other processes. Because of its interpretation, its probability density function should have a subset of $\mathbb{R}^+$ as support. We have chosen the exponential distribution:
\begin{equation} \label{eq:xi}
	\rho( \xi_i ) = \lambda_i e^{-\lambda_i \xi_i} \, ,
\end{equation}
where $\lambda_i$ can be interpreted as the inverse of the mean value of the spontaneous losses that can be produced in the $i$-th process. The exponential distribution has the property that about $63\%$ of its random extractions are lower than its mean value. This means that only few of the extraction will be able to exceed the threshold (unless it is chosen to be less than the mean value). This behavior seems to capture the intuitive picture of a spontaneous loss, which is something that is rarely expected to happen.

We point out that in the context of OR the crucial quantity is the cumulative loss up to the time $t$:
\begin{equation} \label{eq:zeta}
	z_i(t) = \sum_{s \leq t} l_i(s) \, ,
\end{equation}
which can be taken as an indicator of how much money the bank has to put aside to face OR over a time horizon $t$. Let us note that such a time horizon is expressed here in units of time steps, meaning that, supposing that one is interested in the distribution of the cumulative losses registered in one year, $t$ in eq.\ \ref{eq:zeta} is equal to one year divided by the length of a time step.

\section{Parameters estimation} \label{sec:parameters_est}
We will show how the parameters $\theta_i$ and $J_{ij}$ can be learned from a database of historical operational losses that keeps track of the amount of each loss together with the time at which and the process in which each loss occurred. In this context, we interpret the database as a realization of the eq.\ \ref{eq:motion} for a number of time steps consistent with the database at our disposal: $t = 0$ is identified with the time at which the oldest loss in the database occurred and $t = T$ with the time at which the newest loss occurred. The length of a time step of the model is therefore the inverse of the temporal resolution of the database of operational losses from which the parameters are estimated. Since there is no risk of ambiguity in this section, we will use the notation $l_i(t)$ to refer to the amount of the loss registered at the time step $t$ in the $i$-th process in the database at our disposal.

In order to estimate $\theta_i$ we look at those events such that $C_{ij}(t) = 0$, $\forall \, j$; for such events eq.\ \ref{eq:motion} reads:
\begin{equation} \label{eq:theta_1}
	l_i(t) = \text{Ramp} \left[ \theta_i + \xi_i(t) \right];
\end{equation}
and the probability that $l_i(t) = 0$ is:
\begin{equation} \label{eq:theta_2}
	\begin{split}
		\Pr \left[l_i(t)=0 \, | \, C_{ij}(t) = 0, \; \forall \, j \right] & = \Pr \left[ \text{Ramp}[\theta_i + \xi_i(t)] = 0\right] \\
		& = \Pr \left[ \theta_i + \xi_i \leq 0 \right] \\
		& = \Pr \left[ \xi_i \leq - \theta_i \right] \, .
	\end{split}
\end{equation}
In order to estimate the left hand side of eq.\ \ref{eq:theta_2} one would need a sample of values of $l_i(t)$, which is not our case, since a database of operational losses provides a unique value for the amount of the loss in the $i$-th process and at the time step $t$. On the other hand, since the distribution of the noise does not depend on time, the right-hand side of eq.\ \ref{eq:theta_2} also does not depend on time and it is reasonable to make the following identification:
\begin{equation} \label{eq:theta_3}
	\begin{split}
		\Pr \left[ l_i=0 \, | \, C_{ij} = 0, \; \forall \, j \right] & = \Pr \left[ \xi_i \leq - \theta_i \right] \\
		& = \int_{0}^{-\theta_i} {\lambda_i e^{-\lambda_i \xi_i} d\xi_i} \\
		& = 1 - e^{\lambda_i \theta_i} \, ,
	\end{split}
\end{equation}
where the left-hand side has the meaning of a frequentist estimate from the database at our disposal:
\begin{equation} \label{eq:theta_4}
	\Pr \left[ l_i=0 \, | \, C_{ij} = 0, \; \forall \, j \right] = \frac{\text{Fr} \left[ (l_i=0), \; (C_{ij} = 0, \; \forall \, j) \right]}{\text{Fr} \left[ C_{ij} = 0, \; \forall \, j \right]}
\end{equation}
i.e. the number of times such that $l_i(t) = 0$ and $C_{ij}(t) = 0$, $\forall \, j$ divided by the number of times such that times such that $C_{ij}(t) = 0$, $\forall \, j$. Dropping the dependence on time from the left-hand side of eq.\ \ref{eq:theta_3} can be interpreted as the assumption that a single trajectory of the system contains all the information needed to perform a reliable estimation of $\theta_i$. Inverting eq.\ \ref{eq:theta_3} we have:
\begin{equation} \label{eq:theta_5}
	\theta_i = \frac{1}{\lambda_i} \log \left( 1 - \Pr \left[ l_i=0 \, | \, C_{ij} = 0, \; \forall \, j \right] \right) \, .
\end{equation}
We note that the values of $\theta_i$ estimated with such a procedure are necessarily smaller than zero.

The estimation of $J_{ij}$ is analogous to the estimation of $\theta_i$, but it is based on a different class of events. Let us look at those events such that $C_{ij}(t) = c$ and $C_{ik}(t) = 0$ for $c = 1, \ldots, t_{ij}^*$ and $k \neq j$; for such events eq.\ \ref{eq:motion} reads:
\begin{equation} \label{eq:gei_1}
	l_i(t) = \text{Ramp} \left[ c J_{ij} + \theta_i + \xi_i(t) \right];
\end{equation}
and the probability that $l_i(t) = 0$ is given by:
\begin{equation} \label{eq:theta_2}
	\begin{split}
		\Pr \left[ l_i(t)=0 \, | \, C_{ij}(t) = c, \, C_{ik}(t) = 0, \; k \neq j \right] & = \Pr \left[ \text{Ramp}[c J_{ij} + \theta_i + \xi_i(t)] = 0\right] \\
		& = \Pr \left[ c J_{ij} + \theta_i + \xi_i(t) \leq 0 \right] \\
		& = \Pr \left[ \xi_i \leq - c J_{ij} - \theta_i \right] \, .\\
	\end{split}
\end{equation}
Making a similar identification to that of eq.\ \ref{eq:theta_3} we have:
\begin{equation} \label{eq:gei_3}
	\begin{split}
		\Pr \left[ l_i=0 \, | \, C_{ij} = c, \, C_{ik} = 0, \; k \neq j \right] & = \Pr \left[ \xi_i \leq - \theta_i - c J_{ij} \right] \\
		& = \int_{0}^{-\theta_i - c J_{ij}} {\lambda_i e^{-\lambda_i \xi_i} d\xi_i} \\
		& = 1 - e^{\lambda_i \left( \theta_i + c J_{ij} \right)} \, ,
	\end{split}
\end{equation}
where, once again, the left-hand side is a frequentist estimate from the database at our disposal:
\begin{equation} \label{eq:gei_4}
	\Pr \left[ l_i=0 \, | \, C_{ij} = c, \, C_{ik} = 0, \; k \neq j \right] = \frac{\text{Fr} \left[ (l_i=0), \; (C_{ij} = c, \, C_{ik} = 0, \; k \neq j) \right]}{\text{Fr} \left[ C_{ij} = c, \, C_{ik} = 0, \; k \neq j \right]}
\end{equation}
and eq.\ \ref{eq:gei_3} can be inverted to obtain:
\begin{equation} \label{eq:gei_5}
	J_{ij} = \frac{1}{c} \left[ - \theta_i + \frac{1}{\lambda_i} \log \left( 1 - \Pr \left[ l_i=0 \, | \, C_{ij} = c, \, C_{ik} = 0, \; k \neq j \right] \right) \right] \, .
\end{equation}
We note from eq.\ \ref{eq:gei_5} that, depending on the class of events found on the database at our disposal, there may be up to $t_{ij}^*$ different estimates of $J_{ij}$: one for each possible value of $c$. The problem of dealing with multiple estimates of $J_{ij}$ should be faced depending on the use one has for the value of the parameter $J_{ij}$. In sections \ref{sec:validation} and \ref{sec:results} we describe two different strategies for collapsing the multiple estimates of $J_{ij}$ into a single value. Let us note that eqs.\ \ref{eq:theta_3} and \ref{eq:gei_3} can be easily generalized for every distribution of $\xi_i(t)$, since $\Pr[\xi_i(t) \leq x]$ is simple the cumulative function calculated in $x$. However, the passage from eqs.\ \ref{eq:theta_3} and \ref{eq:gei_3} to eqs.\ \ref{eq:theta_5} and \ref{eq:gei_5} is possible only if such a cumulative function is invertible. In all the other cases eqs.\ \ref{eq:theta_3} and \ref{eq:gei_3} must be solved numerically to obtain the estimates of $\theta_i$ and $J_{ij}$.

With regard to the estimation of $\lambda_i$, a trivial possibility is to invert eq.\ \ref{eq:theta_5}, provided that the value of $\theta_i$ is known. Let us recall that $\theta_i$ can be interpreted as the money invested by the bank on the $i$-th process to keep it working properly and, therefore, it is plausible that some kind of knowledge about its value is available. Nevertheless, $\theta_i$ may also be unknown, since, in general, it is the threshold that a potential loss in the $i$-th process has to overcome to become an actual loss. The value of $\lambda_i$ may be assessed independently from $\theta_i$ only if some information regarding the spontaneous losses in the $i$-th process is available. Sometimes one may lack the knowledge regarding the interactions of a particular process with the others, but have a rather precise idea regarding the distribution of spontaneous losses instead. In those cases, we see from eq.\ \ref{eq:xi} that $\lambda_i$ is the inverse of the mean of the spontaneous losses in the $i$-th process or it can obtained from any quantile $q_i$ of order $\alpha$ of the distribution of spontaneous losses of the $i$-th process: $\lambda_i = - \log(1 - \alpha_i)/q_i$.

Furthermore $t_{ij}^*$ cannot be learned directly from a database of operational losses and should be assessed from domain experts instead. Recalling the definition of $t_{ij}^*$ given in section \ref{sec:model}, the question that should be answered in order to assess its value is: ``which is the maximum delay up to which the losses occurred in the $j$-th process may influence the losses in the $i$-th process?''.

Let us for a moment comment on the total number of parameters of the model that must be estimated or assessed. There are $N^2$ couplings $J_{ij}$, $N^2$ maximum correlation times $t_{ij}^*$ since they both depend on two indexes, $N$ supports $\theta_i$ and $N$ inverse means of the noise $\lambda_i$ since they depend only on one index, so that the total number of parameters is $2(N^2 + N)$. In the context of the LDA, supposing that the frequencies and severities of the all processes are independent, one has that the total number of parameters is $(n_f + n_s) N$, where $n_f$ and $n_s$ are the number of parameters of the chosen frequency and severity distribution, respectively. Typically $n_f$ = 1, 2 (for Poisson and negative binomial distributions) and $n_s$ = 2 (for lognormal, gamma or Weibull distributions); obviously $n_s$ becomes larger if the severity tails are fitted with the extreme value theory. However, one should consider that the proposed approach takes the correlations among the different processes into account. For this reason it is fair to compare it with the LDA using couplae to capture the correlations. In this case (apart from the trivial cases of comonotone or anti-conomotone copulae) one has to use one copula for each couple of processes, resulting in additional $n_c N^2$ parameters, where $n_c$ is the number of parameters of each copula. Typically $n_c$ = 1 (for Gaussian, Clayton, Gumbel or Frank copula). The total number of parameters of the LDA with copulas is then $n_c N^2 + (n_f + n_s)N$, which is of the same order in $N$ as the proposed approach.

\section{Validation} \label{sec:validation}
In order to validate the proposed procedure for estimating the parameters we propose the following steps:
\begin{enumerate}
	\item set the parameters $\theta_i$, $J_{ij}$, $\lambda_i$ and $t_{ij}^*$ of the model to realistic values. Let eq.\ \ref{eq:motion} evolve for $T$ time steps and consider the obtained values of $l_i(t)$ to be a database of operational losses.
	\item Estimate the parameters $\theta_i$ and $J_{ij}$ using the procedure proposed in section \ref{sec:parameters_est} and compare the obtained values with the ones set out in point 1.
	\item Simulate eq.\ \ref{eq:motion} a large number of times using the estimated parameters, so that a sample of trajectories of the system is obtained. Since there may be more than one estimate for each $J_{ij}$, the values of their estimates should be sampled among the available ones for each simulated trajectory.
	\item Compare $z_i(t)$ calculated from the database generated in point 1 with the average of the same quantity calculated from the sample of trajectories generated in point 3. There are two reasons not to use $l_i(t)$ and its average: on one hand, the single realization of eq.\ \ref{eq:motion} strongly depends on the extractions of the noise and thus is not directly comparable to another realization; on the other hand, as previously mentioned, the quantity of interest in OR is not the amount of money lost at a certain time step, but rather the total amount of money lost up to that certain time, which is precisely the meaning of $z_i(t)$.
\end{enumerate}
It is possible to look not only at the sample average of $z_i(t)$, but also at the full distribution of the values of $z_i(t)$. In particular, it is possible to estimate the VaR at a given time step by numerically calculating some percentile of the sampled distribution at the desired time step. The forecasting power of the model can be tested by repeating points from 2 to 5 using only a fraction of the database generated in point 1 to estimate the parameters $\theta_i$ and $J_{ij}$, but still simulating eq.\ \ref{eq:motion} in point 3 for $T$ time steps. In this way we are effectively ignoring the information contained in the neglected fraction of the original database and making a forecast relative to those time steps.

\section{Chosen parameters} \label{sec:parameters_ch}
We simulate for $T = 200000$ time steps a system composed by $N = 5$ processes whose parameters are set up so that they can mimic the interactions between the following realistic processes:
\begin{enumerate}
	\item machine failure,
	\item human error,
	\item internal fraud,
	\item failed transaction of type I,
	\item failed transaction of type II.
\end{enumerate}
Before we start to justify the chosen parameters, let us make a couple of observations. The value of $\theta_i$ may be chosen to fix the unit of measurement of $l_i(t)$. In fact, rescaling all the terms in eq.\ \ref{eq:motion} by a factor $\theta_i$, we obtain:
\begin{equation}
	\frac{l_i(t)}{\theta_i} = \text{Ramp} \left( \sum_{j=1}^N \frac{J_{ij}}{\theta_i} C_{ij}(t) + 1 + \frac{\xi_i(t)}{\theta_i} \right) \, ,
\end{equation}
which is precisely the same equation as eq.\ \ref{eq:motion}, but with rescaled losses and parameters. This means that, since we are free to choose the unit of measure for $l_i(t)$, it is perfectly legitimate to fix $\theta_i = 1$. However, since we want $\theta_i$ to model the effort made by the bank to avoid losses, as we already pointed out, its value must be negative, thus:
\begin{equation*}
	\vec{\theta} = (-1, -1, -1, -1, -1) \, .
\end{equation*}

Rather than directly specifying the value of $\lambda_i$, it is possible to specify the probability $p_i$ that a loss occurs in a noninteracting process. $p_i$ should be easier to assess, since it does not carry any information about the interactions. It is proportional to the number of times that a loss would occur in a given process if the process is left ``all alone'', with no interaction with the other processes. Dropping the interaction term from eq.\ \ref{eq:motion} and calculating $p_i$ one obtains:
\begin{equation}
	\lambda_i = \frac{\log p_i}{\theta_i}
\end{equation}
that links $\lambda_i$ to $p_i$. The chosen values for $p_i$ are:
\begin{equation*}
	\vec{p} = (0.01, 0.05, 0.01, 0.025, 0.025)
\end{equation*}
which are consistent with the fact that the probability that a human error spontaneously produces a loss is much higher (5 times) than the same probability for a machinery fault. $p_3$ is equally low, assuming that a loss generated by a spontaneous internal fraud is as rare as a spontaneous loss generated by a machinery failure, while intermediate values are chosen for failed transactions.

The $J_{ij}$ are all equal to zero, apart from:
\begin{equation*}
	J_{12} = 0.1
\end{equation*}
which accounts for the possibility that a human error causes a machine failure,
\begin{equation*}
	J_{33} = 0.15
\end{equation*}
which accounts for the possibility that the act of committing a fraud leads to committing fraud again,
\begin{equation*}
	J_{43} = J_{53}= 0.15
\end{equation*}
which accounts for the possibility that some transactions may fail because some funds have been subtracted by a fraud. The two types of transaction failures are distinguished by the fact that they may be also influenced by different processes, that is, type I by a human error and type II by a machine failure:
\begin{equation*}
	J_{42} = J_{51}= 0.1 \, ,
\end{equation*}
but in both cases the consequences are minor with respect to those deriving from a fund subtraction due to an internal fraud.

From eq.\ \ref{eq:motion} we see that the only relevant values of $t_{ij}^*$	are those relative to the values of $J_{ij}$ which are different from zero. In all those cases we set $t_{ij}^* = 5$, which takes into account the possibility of different-time correlations. 

We conclude this section by pointing out that eq.\ \ref{eq:motion} requires a number of time steps in the initial condition which is equal to $\max_{i,j} t_{ij}^* = 5$. We set $l_i(t) = 0$ for $t = 1, \ldots, 5$, i.e. all process are perfectly working at the beginning, without losses.

\section{Results} \label{sec:results}
Using the parameters described in section \ref{sec:parameters_ch} we have followed the validation protocol and estimated the parameters using the whole database generated at point 1. We repeated the procedure using only the first three-quarters of its time steps to test the forecasting power of the model. Using the full database we find the following relative errors on the estimated parameters:
\begin{gather*}
	\delta \vec{\theta} \simeq (0.01, \,< 0.01, \, 0.01, \, < 0.01, \, < 0.01) \\
	\delta J_{12} \lesssim 0.01  \qquad \delta J_{33} \simeq 0.05 \qquad \delta J_{42} \simeq 0.02 \\
	\delta J_{43} \simeq 0.03  \qquad \delta J_{51} \simeq 0.08 \qquad \delta J_{53} \simeq 0.06 \, ,
\end{gather*}
while, using only the first three-quarters of its time steps, we find:
\begin{gather*}
	\delta \vec{\theta} \simeq (0.01, \,< 0.01, \, 0.01, \, < 0.01, \, < 0.01) \\
	\delta J_{12} \simeq 0.07  \qquad \delta J_{33} \lesssim 0.01 \qquad \delta J_{42} \simeq 0.01 \\
	\delta J_{43} \simeq 0.02  \qquad \delta J_{51} \simeq 0.02 \qquad \delta J_{53} \simeq 0.04 \, ,
\end{gather*}
where the relative error on $J_{ij}$ has been calculated using the mean of the available estimates. We can immediately note that the errors on the estimated parameters are comparable in the two cases and, in some cases, are lower when only the first three quarters of the available time steps are used. This means that the information contained in last quarter of time steps is redundant and that the first three-quarters of the available time steps can provide all the information needed to perform a reliable estimation of the parameters. This is a very general consideration indeed: every time we want to make a forecast about some quantity we are assuming that our knowledge about its dynamics is complete, i.e. all the relevant information is contained in the past (already observed) evolution.

\begin{figure}[t]
	\centering
	\includegraphics[width=0.48\columnwidth]{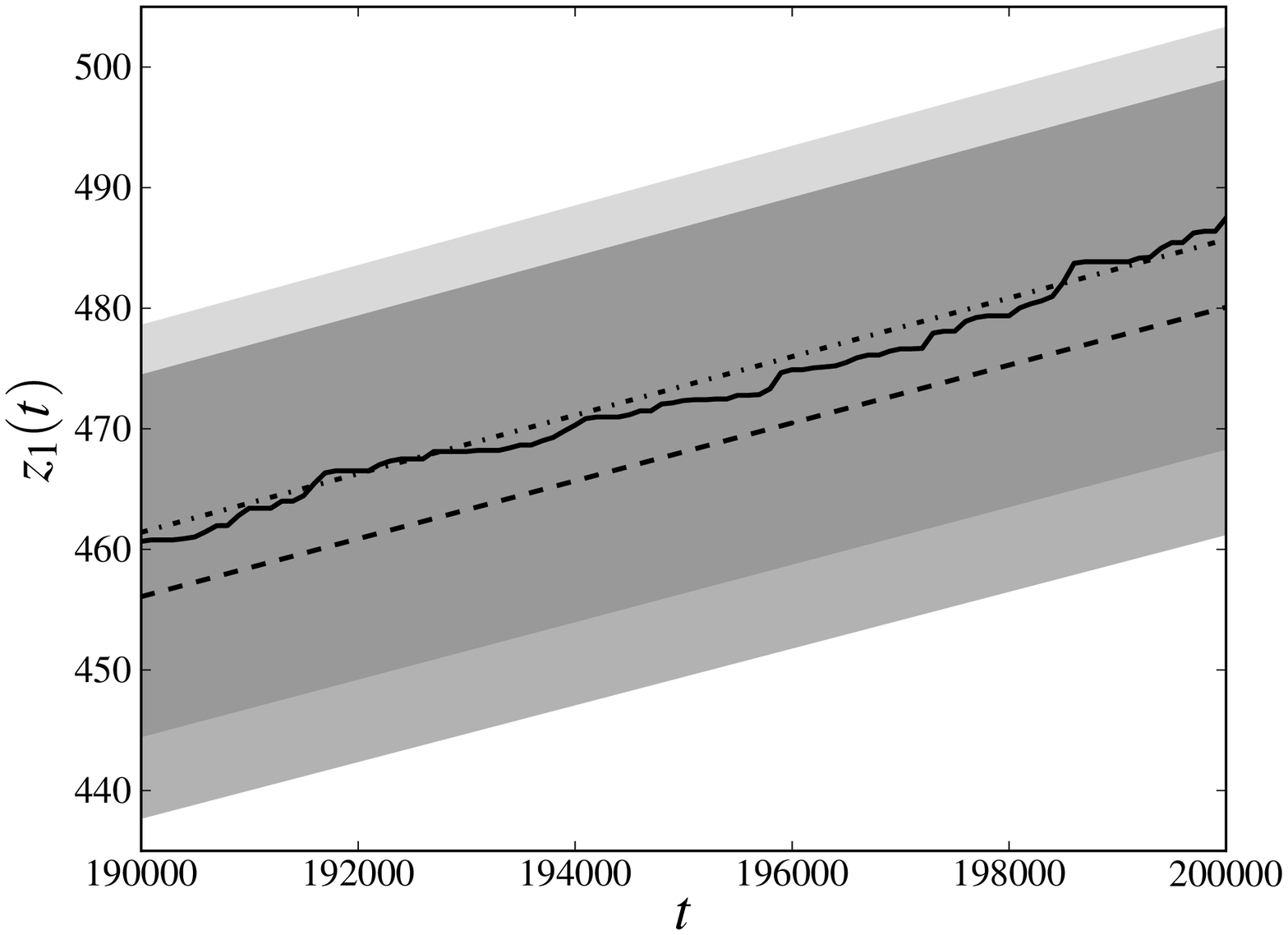}
	\includegraphics[width=0.48\columnwidth]{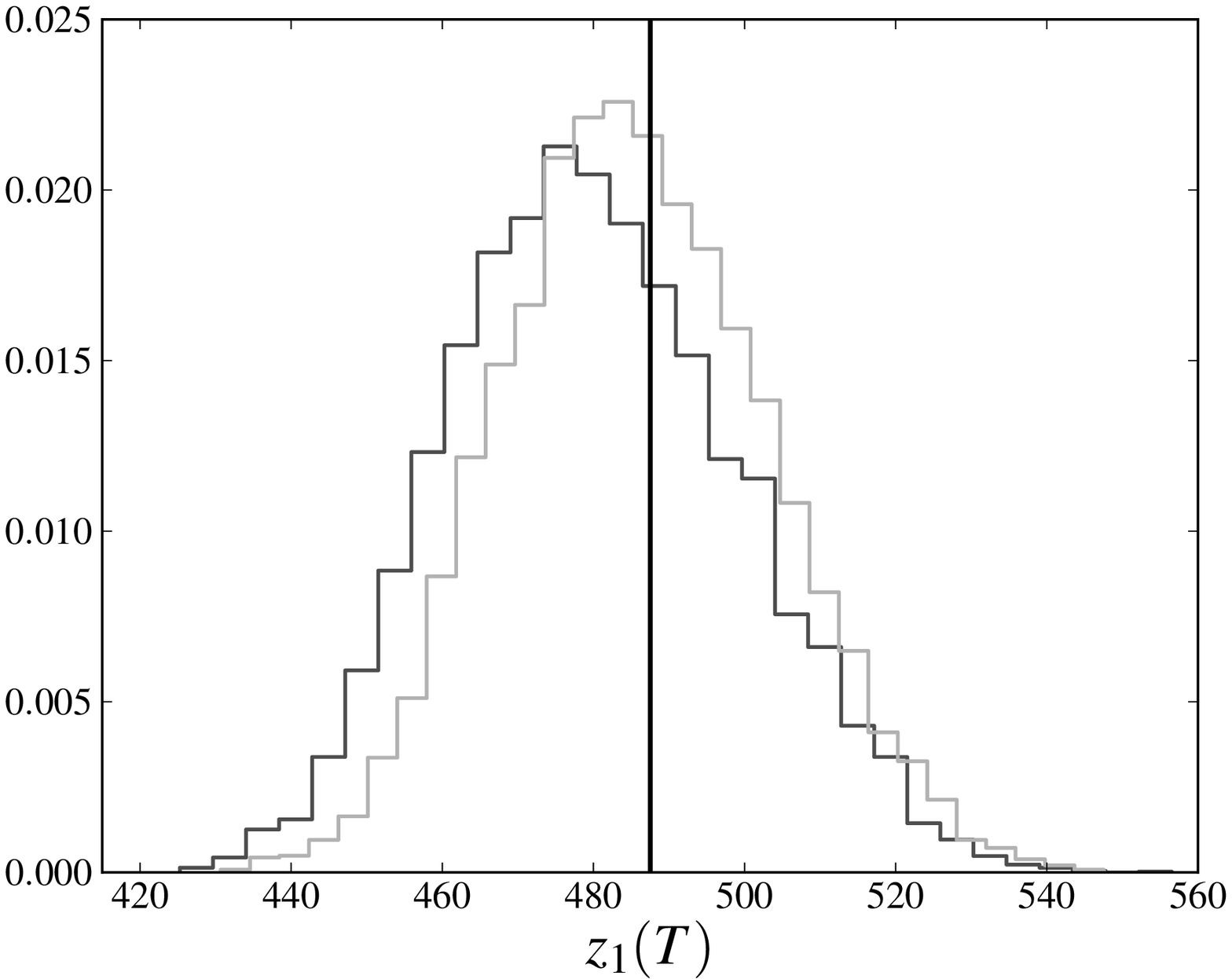}	
	\includegraphics[width=0.48\columnwidth]{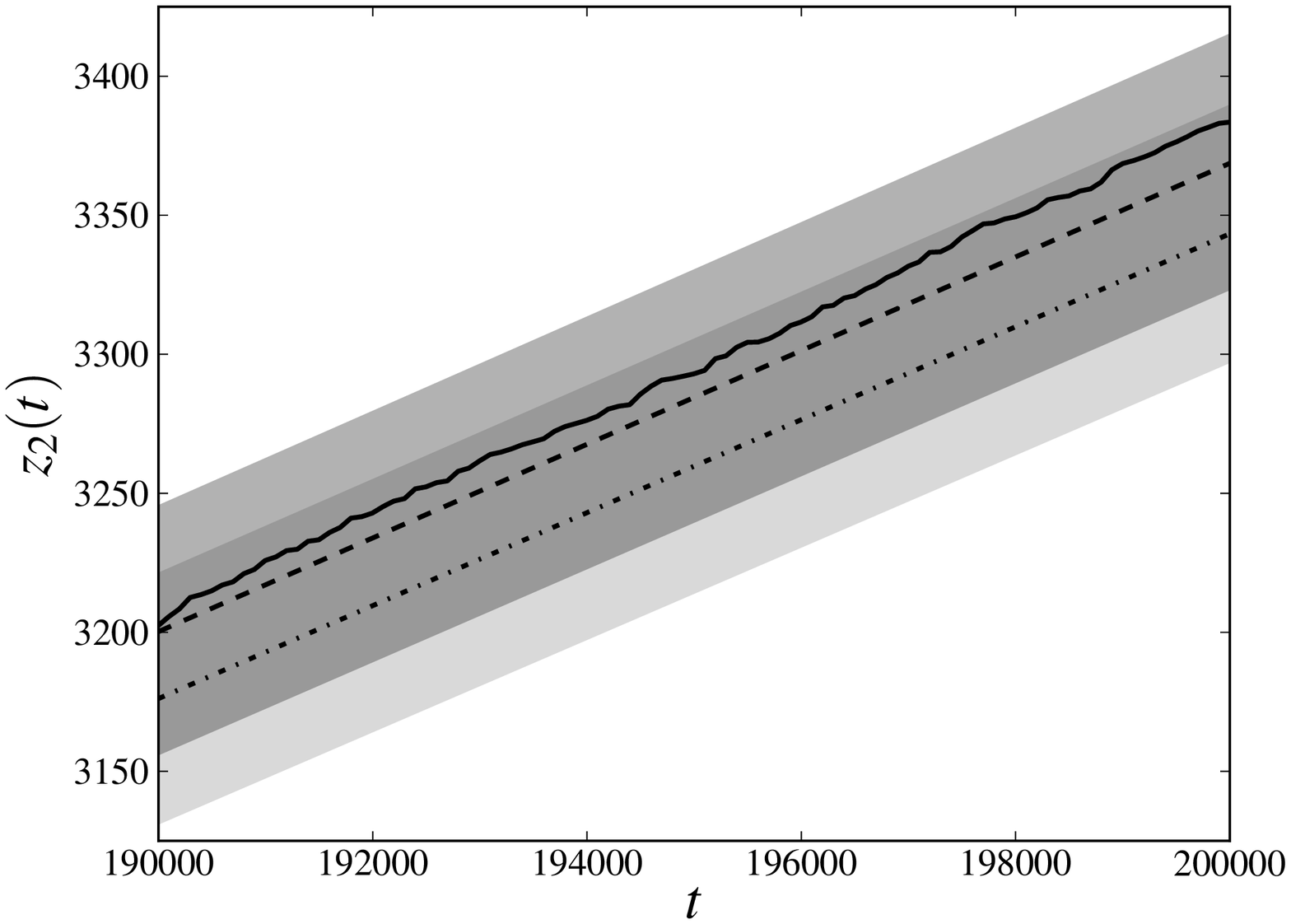}
	\includegraphics[width=0.48\columnwidth]{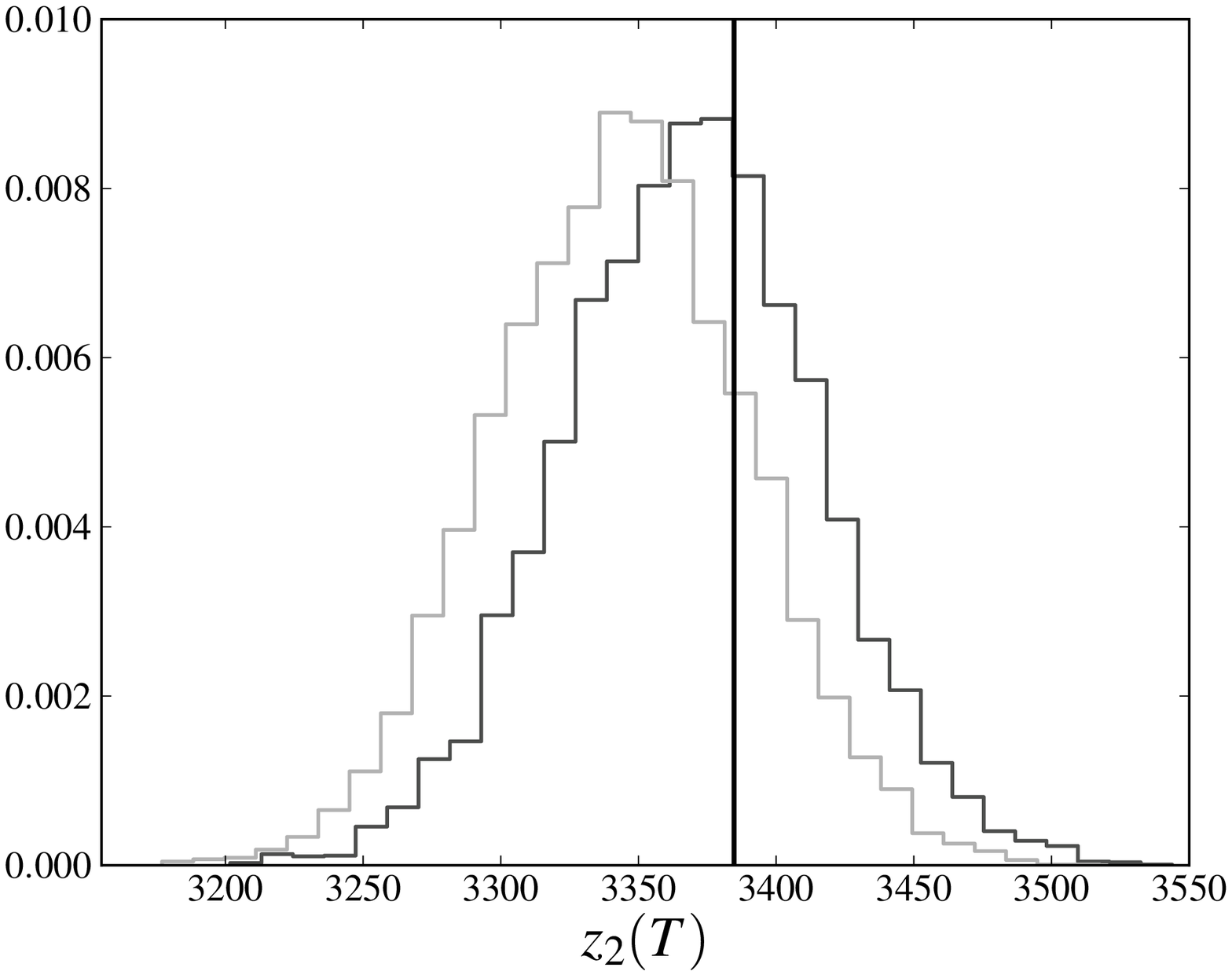}
	\includegraphics[width=0.48\columnwidth]{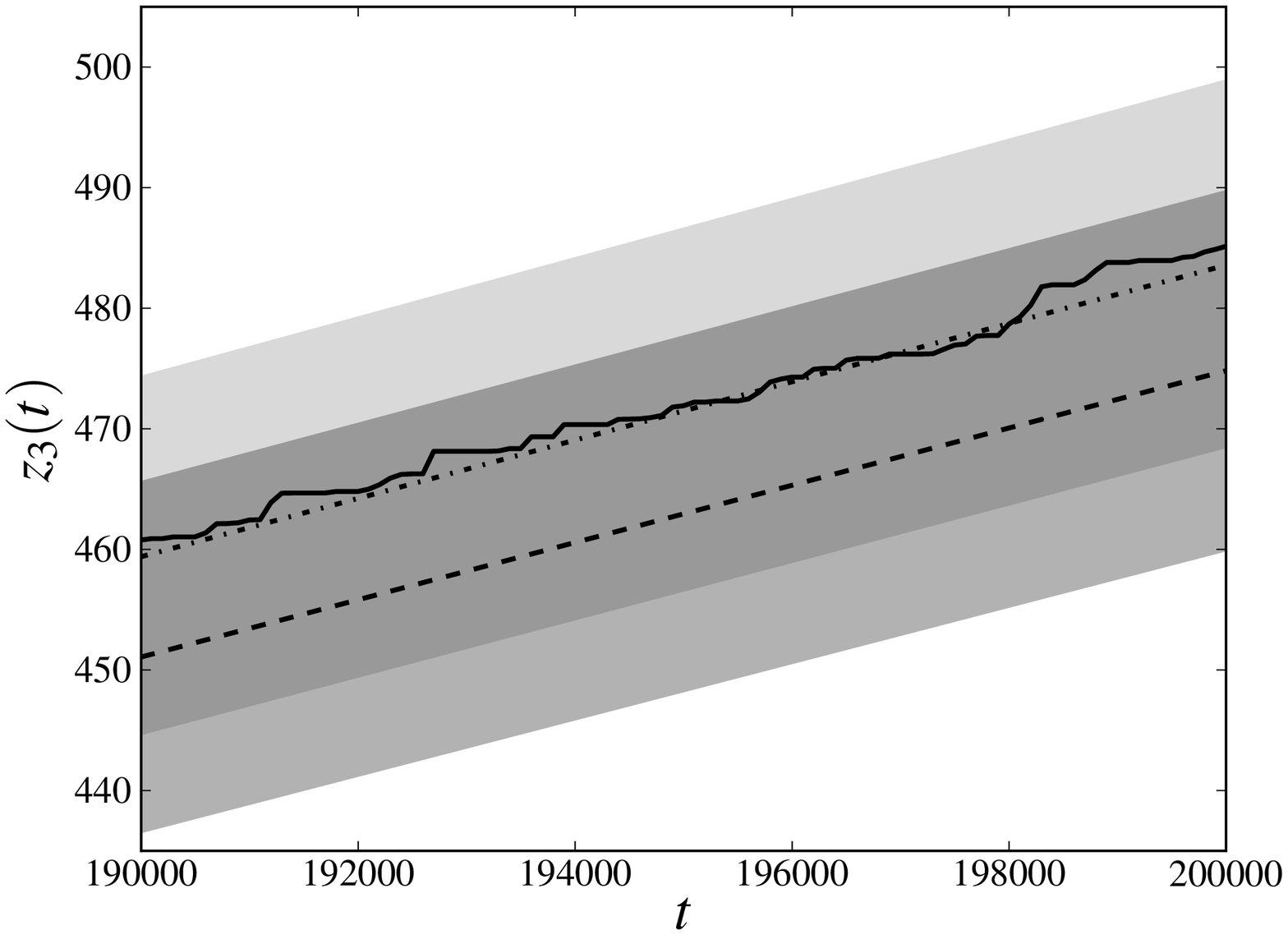}
	\includegraphics[width=0.48\columnwidth]{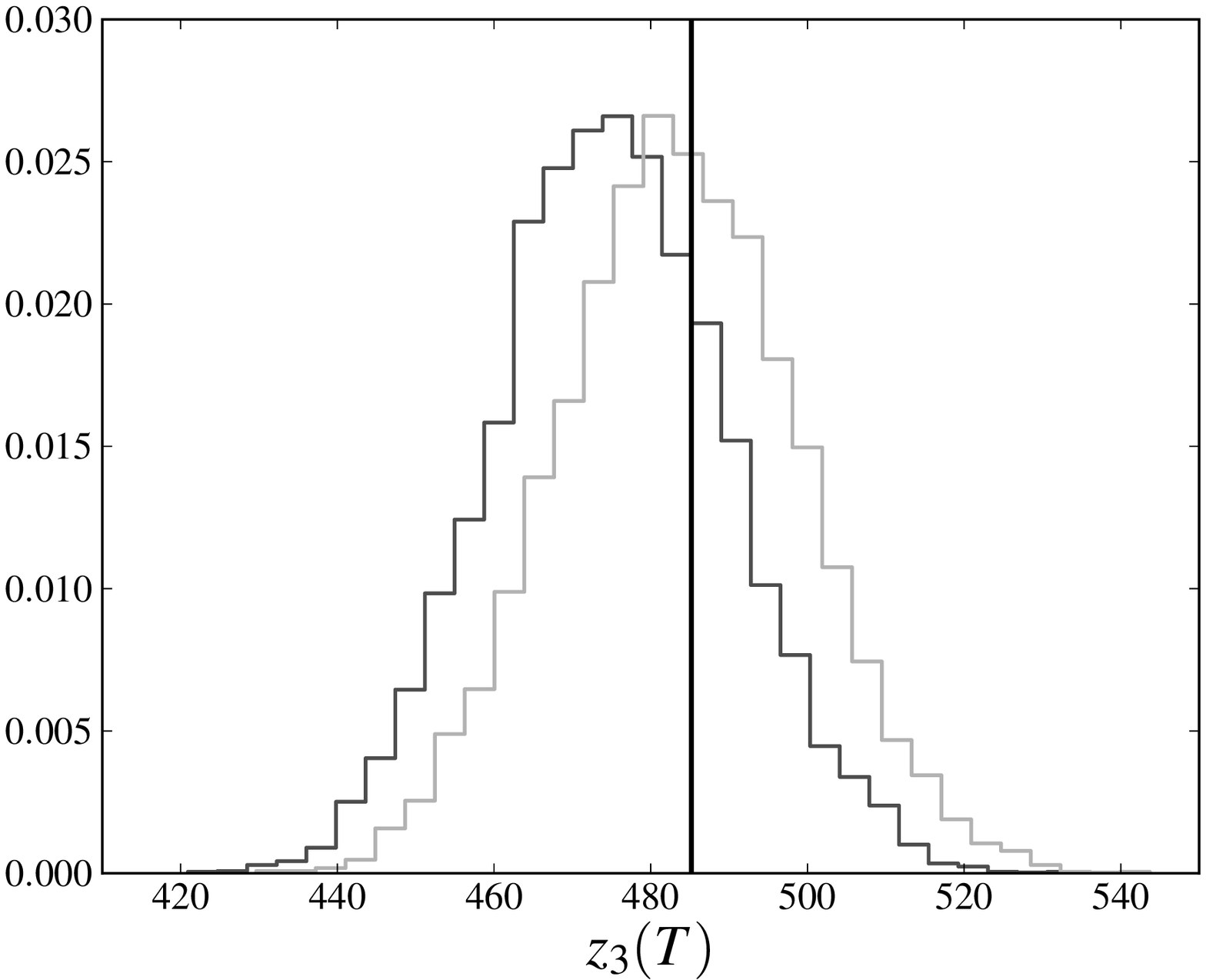}			
	\caption{Left panels: cumulative loss of the original trajectory (solid line) and average of $z_i(t)$ obtained estimating the parameters from the original trajectory (dashed line) and only from three quarters of the available time steps (dash-dotted line); the limits of the semi-transparent regions encompass one standard deviations around the averages. Right panels: cumulative loss of the original trajectory (solid line) and distribution of $z_i(T)$ of the sampled evolutions obtained estimating the parameters from the original trajectory (darker histogram) and only from three quarters of it (lighter histogram). Processes from 1 to 3.}
	\label{fig:part_1}
\end{figure}

\begin{figure}[t]
	\centering
	\includegraphics[width=0.48\columnwidth]{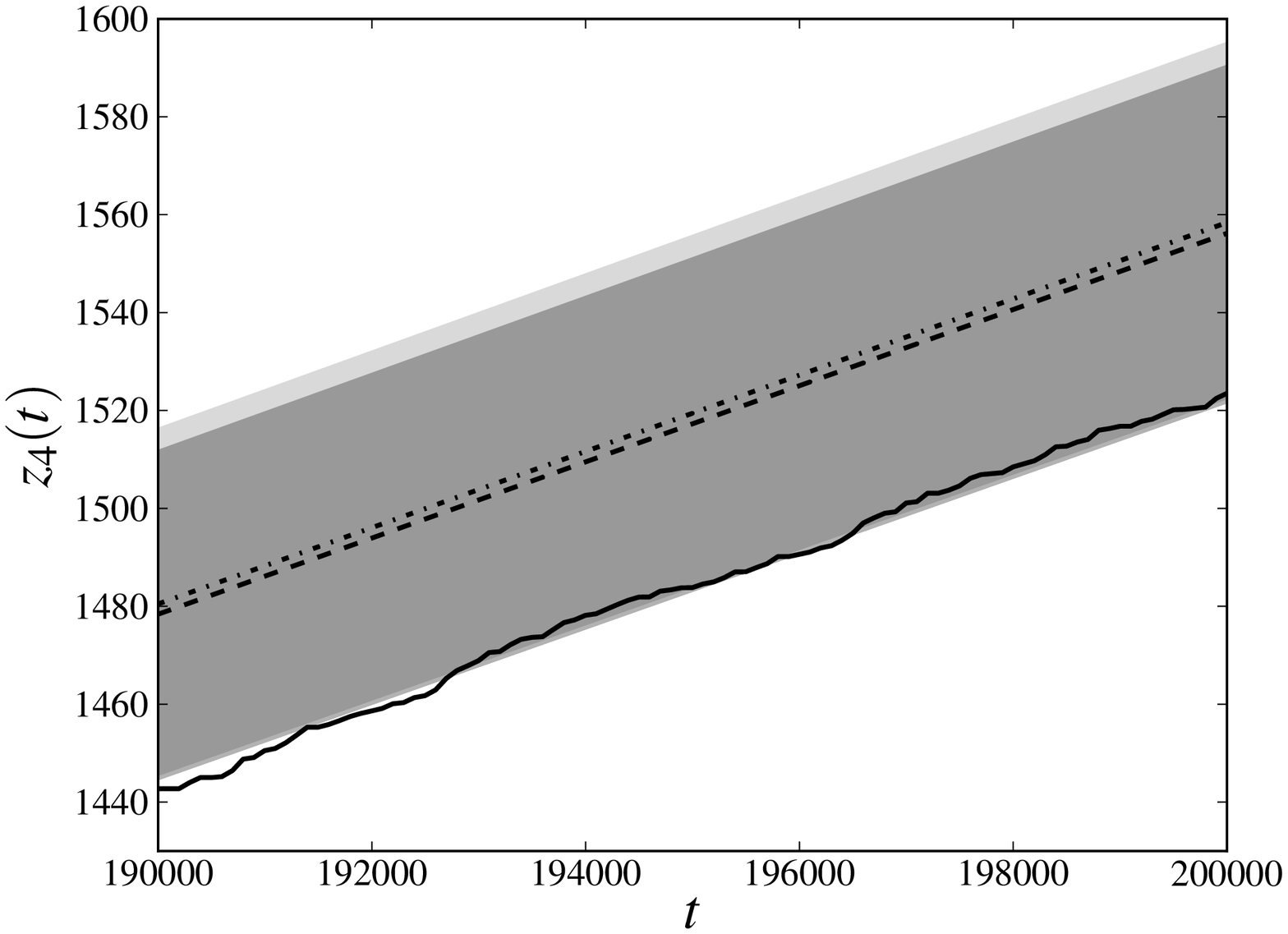}
	\includegraphics[width=0.48\columnwidth]{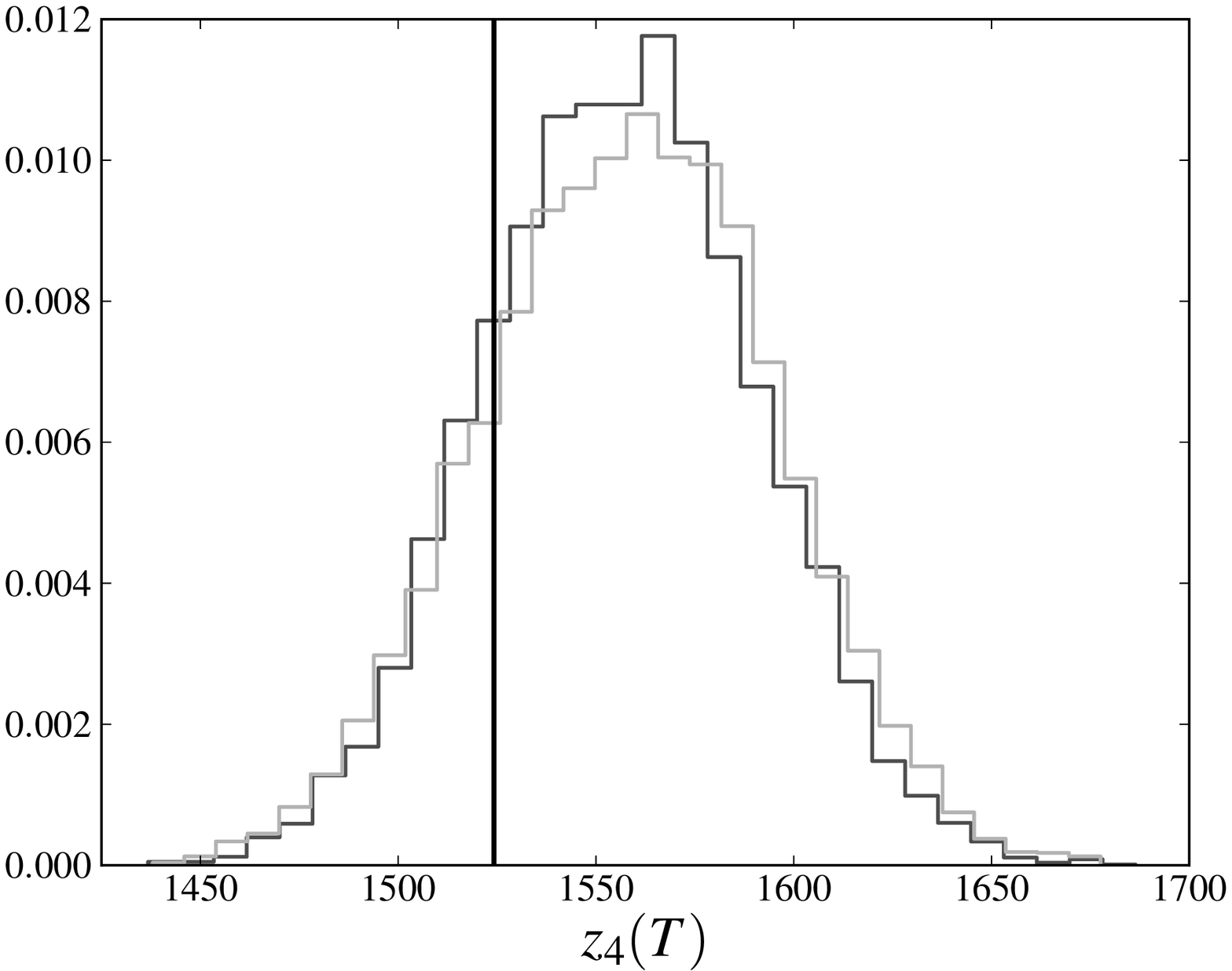}	
	\includegraphics[width=0.48\columnwidth]{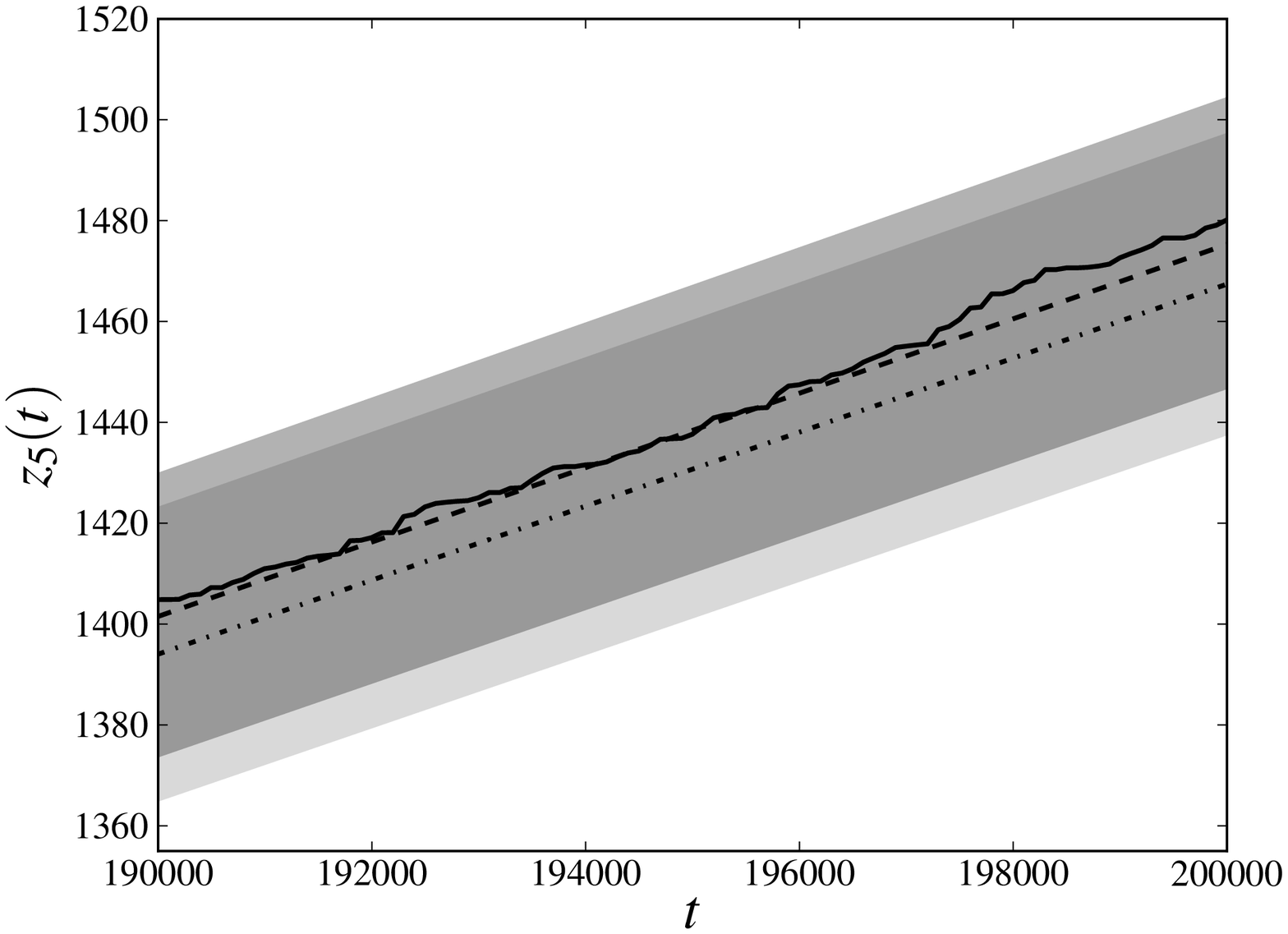}
	\includegraphics[width=0.48\columnwidth]{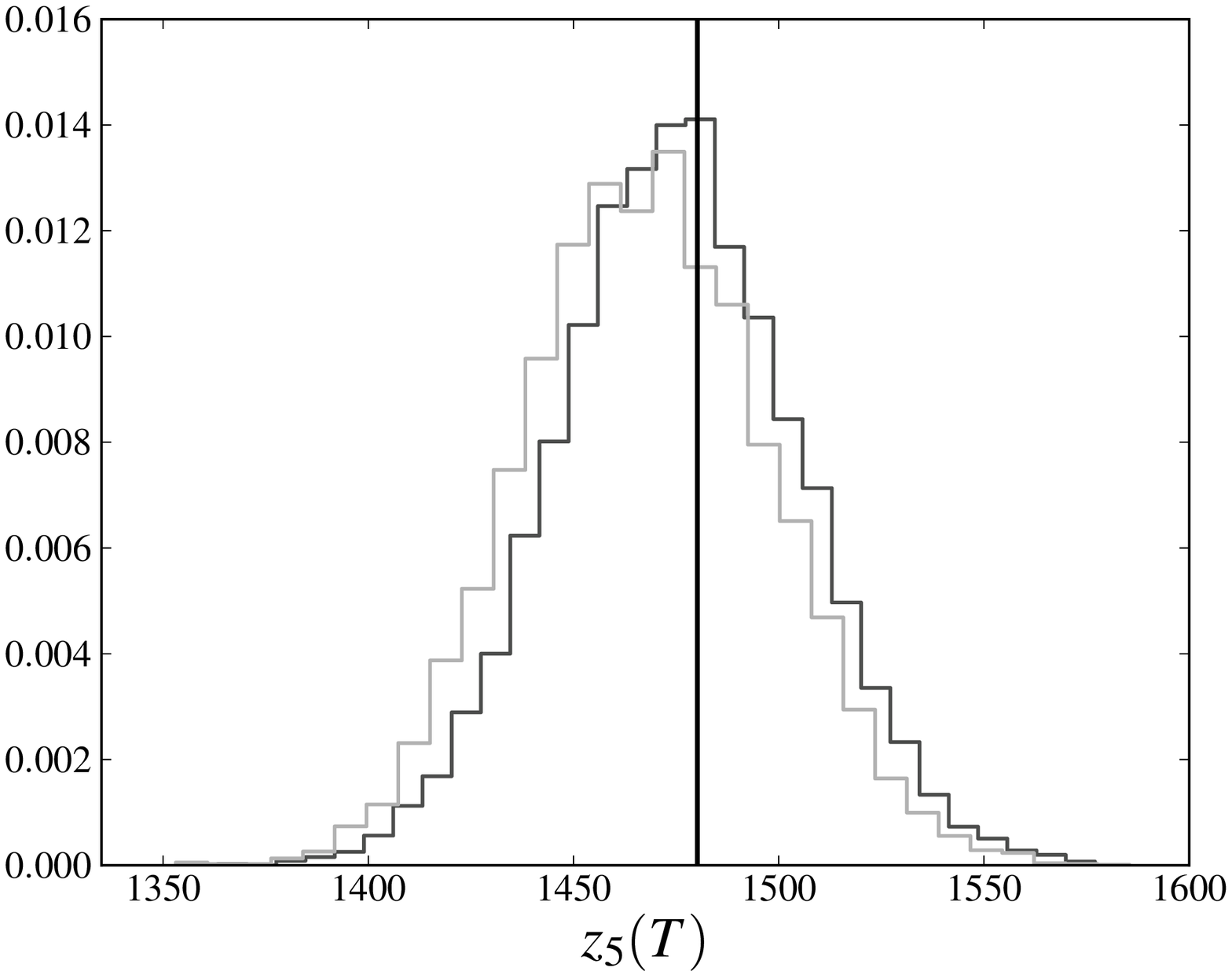}
	\caption{Both right and left panels as in fig.\ \ref{fig:part_1}, but for processes 4 and 5.}
	\label{fig:part_2}
\end{figure}

In the left panels of figs.\ \ref{fig:part_1} and \ref{fig:part_2} we compare $z_i(t)$ (only the last 10000 time steps, for the sake of readability) calculated for the database generated at point 1 with the average of the same quantity for the trajectories sampled at point 3. The solid line denotes $z_i(t)$ relative to the original trajectory, the dashed line denotes the average of $z_i(t)$ obtained estimating the parameters from the whole database and the dash-dotted line denotes the average of $z_i(t)$ obtained estimating the parameters from only the first three-quarters of the available time steps. The darker transparent region spans a standard deviation around the average of $z_i(t)$ obtained estimating the parameters from the whole database. The lighter transparent region has the same meaning for the average of $z_i(t)$ obtained estimating the parameters only from the first three-quarters of the available time steps. We see that, in all the cases, the original trajectory is reproduced with an error which is less than (or, for the $4$-th process, equal to) one standard deviation, meaning that the cumulative loss in the last part of the database has been reliably forecast.

In the right panels of figs.\ \ref{fig:part_1} and \ref{fig:part_2} we compare the full distribution of $z_i(T)$ of the sampled evolutions at point 3 with the original trajectory. The solid line denotes $z_i(T)$ of the original trajectory, the darker histogram refers to the distribution of $z_i(T)$ obtained estimating the parameters from the whole database, while lighter histogram refers to the distribution of $z_i(T)$ obtained estimating the parameters only from the first three-quarters of the available time steps. Again we see that peaks of both distributions are very close and almost coincident with the value of the original trajectory.

We note that the observed average of $z_i(t)$ is, by a good approximation, linear and that the distribution of $z_i(T)$ does not seem to exhibit heavy tails. Such a behavior is also present in the process 2, which is not influenced by the losses incurred in any other process and whose losses can only have been spontaneously generated by the noise term. This observation encourages us to explore models in which the distribution of the noise has a different shape and heavy tails. However, it should be also kept in mind that deviations from the behavior shown here may depend not only on the distribution of the noise, but also on the chosen parameters. As already pointed out in section \ref{sec:intro}, a further exploration of the parameter space is certainly needed to understand if this peculiarity depends on the particular choice of parameters that has been made.

\section{Conclusions}
In this paper we have proposed a new dynamical model for OR. To the best of our knowledge for the first time the loss distribution is derived from an entirely dynamical approach. In a previous paper \citep{kuhn-neu} such an approach was limited to the frequency distribution. The use of a dynamical approach has a great methodological advantage over static approaches. For example, the LDA uses historical data to fit the yearly loss distribution and, exploiting this distribution to estimate the capital charge for the next year, the method implicitly assumes that the distribution will not change from one year to the next. On the other hand, a dynamical approach provides a natural framework for forecasting the distribution of the future losses on which the capital charge requirements should be based, making the much weaker assumption that only the basic mechanisms of loss production do not change from year to year. Indeed, also this assumption may be partially relaxed by estimating the parameters only from the most recent part of a database of operational losses or by forcing them to values assessed by a domain experts. However, we point out that, since the meaning of the parameters of the model is unconventional in the context of OR, such an assessment must be carried out with extreme care, so that the experts can be completely aware of the role played by each parameter in the model. Another crucial feature of the proposed approach that is absent in most of the AMAs is that it can take into account different-time correlations among the processes by means of the interaction term in the equation of motion.

Let us remark that the current implementation adheres to many of the AMA guidelines: i) some parameters of the model are estimated from a database of internal operational losses, ii) the remaining parameters have to be assessed by domain experts, iii) the processes can be considered to be, or can be linked to, the 56 (business lines, event types) couples, iv) the VaR of the cumulative loss distribution at every time step can be calculated with the desired level of confidence; the calculation of the VaR over the time horizon of one year can be performed once the fictional time scale the model is linked to some real time scale, which may be the time resolution of the available database of operational losses. It is worth noting that the proposed model does not require massive investments for its implementation. In fact, only a reliable monitoring of the internal losses and the assessments of internal experts are needed. Both of these things are requirements that every AMA-oriented bank should meet in any case. From a practical point of view the main steps required to implement the proposed approach are the following: i) the processes should be identified, since (as hinted in \ref{sec:intro}) they may strongly depend on the specific bank; ii) the losses incurred in the processes have to be monitored for a sufficient amount of time, so that a reliable database of operational losses can be built and iii) the parameters that cannot be estimated from the database must be assessed by domain experts.

The current limitations of the model pave the way for future research directions. A recent paper \citep{bardoscia-bellotti} has focused on the case in which no causal loops among the processes exist, showing that, together with $\theta_i$ and $J_{ij}$, even the parameter $\lambda_i$ can be learned from a database of operational losses in that case. We point out that the absence of causal loops is an hypothesis that is often accepted. In particular this is the crucial hypothesis of all the approaches based on bayesian networks \citep{cowell-verral-yoon}. In this case the loss distribution has been analytically derived as well, establishing a deep connection between the properties of the noise in the equation of motion and the shape of the loss distribution. We have already discussed the importance of investigating the case in which the distribution of the noise has heavy tails. \citet{bardoscia} deals with such distributions, generalizing the results obtained in \citet{bardoscia-bellotti} and proving that, at least in the case in which there are no causal loops, the distribution of the cumulative loss is heavy tailed if and only if the distribution of the noise is heavy tailed. Further research should be certainly devoted to investigating the more general case in which causal loops are present.

\section*{Acknowledgments}
M.~B. would like to thank Maria Valentina Carlucci for the countless suggestions and fruitful discussions.

\bibliographystyle{chicago}

\end{document}